\documentclass[letterpaper,prc,twocolumn,showpacs,floatfix,nofootinbib,preprintnumbers,superscriptaddress,amsmath,amssymb]{revtex4-1} 
\usepackage{graphicx}
\usepackage{bm}
\usepackage{wrapfig}
\usepackage{amsmath,amssymb}
\usepackage{color}

\newcommand{\bra}[1]{\langle {#1} |}
\newcommand{\ket}[1]{| {#1} \rangle}

\def\vect#1{{\mbox{\boldmath $#1$}}}

\begin{document}

% Use the \preprint command to place your local institutional report
% number in the upper righthand corner of the title page in preprint mode.
% Multiple \preprint commands are allowed.
% Use the 'preprintnumbers' class option to override journal defaults
% to display numbers if necessary
%\preprint{}

%Title of paper
\title{Low-lying 2$^+$ states generated by $pn$-quadrupole correlation and $N$=28 shell quenching}

% repeat the \author .. \affiliation  etc. as needed
% \email, \thanks, \homepage, \altaffiliation all apply to the current
% author. Explanatory text should go in the []'s, actual e-mail
% address or url should go in the {}'s for \email and \homepage.
% Please use the appropriate macro foreach each type of information

% \affiliation command applies to all authors since the last
% \affiliation command. The \affiliation command should follow the
% other information
% \affiliation can be followed by \email, \homepage, \thanks as well.
\author{Shuichiro~Ebata}
\affiliation{Meme Media Laboratory, Hokkaido University, Sapporo, 060-0813, Japan}
\author{Masaaki~Kimura}
\affiliation{Department of Physics, Hokkaido University, Sapporo, 060-0810, Japan}
% \author{Takashi~Nakatsukasa}
% \affiliation{RIKEN Nishina Center, Wako-shi 351-0198, Japan}
% \affiliation{Center for Computational Sciences, University of Tsukuba, Tsukuba 305-8571, Japan}
%\email[]{Your e-mail address}
%\homepage[]{Your web page}
%\thanks{}
%\altaffiliation{}
%\affiliation{}

%Collaboration name if desired (requires use of superscriptaddress
%option in \documentclass). \noaffiliation is required (may also be
%used with the \author command).
%\collaboration can be followed by \email, \homepage, \thanks as well.
%\collaboration{}
%\noaffiliation

\date{\today}

\begin{abstract}
The quadrupole vibrational modes of neutron-rich $N$=28 isotones ($^{48}$Ca, $^{46}$Ar,
$^{44}$S and $^{42}$Si) are investigated  using the canonical-basis time-dependent
Hartree-Fock-Bogoliubov theory with several choice of energy density functionals, 
including nuclear pairing correlation. 
It is found that the quenching of $N$=28 shell gap and the proton holes 
in the $sd$-shell trigger quadrupole correlation and 
increase the collectivity of the low-lying 2$^+$ state in $^{46}$Ar. 
It is also found that the pairing correlation plays an important role 
to increase the collectivity. 
We also demonstrate that the same mechanism to enhance the low-lying collectivity applies 
to other $N$=28 isotones $^{44}$S and $^{42}$Si, and 
it generates a couple of low-lying 2$^+$ states which can be associated with the observed $2^+$ states. 
\end{abstract}

% insert suggested PACS numbers in braces on next line
% insert suggested keywords - APS authors don't need to do this
%\keywords{}

%\maketitle must follow title, authors, abstract, \pacs, and \keywords
\maketitle
% body of paper here - Use proper section commands
% References should be done using the \cite, \ref, and \label commands
%\section{}
% Put \label in argument of \section for cross-referencing
%\section{\label{}}
%\subsection{}
%\subsubsection{}

\section{Introduction}
\label{sec:introduction}
As represented by the pygmy dipole mode \cite{SIS90, aL01, pA05, jG08}, 
there are novel excitation modes peculiar to unstable nuclei. 
In particular, the low-lying collective excitation modes are quite
sensitive to the underlying shell structure and the pairing correlations, and hence,
the quenched shell gaps in unstable nuclei should generate unique collective modes. 

One of the interesting example is the $N$=28 shell gap that is known to be quenched
in the vicinity of $^{44}$S, and has been paid considerable experimental 
\cite{hS96,aG03,laR05,dM10,tG97,fS00,dS02,sG04,jF06,bJ07,mZ09,cF10,dsG11,jF05,bB07,lG08,sT12}
and theoretical attentions \cite{RCF97,djD99,LVR99,PGF00,RER02,eC05,lG10,KSM11,zpLi11,RE11,mK13}.
Since the the orbital angular momenta of $f_{7/2}$ and $p_{3/2}$ differ by two, 
the quench of $N$=28 shell gap will lead to the strong quadrupole correlation of neutrons. 
Furthermore, the protons in Si, S and Ar isotopes occupy the middle of the $sd$-shell, 
and hence, the quadrupole correlation should also
exist in the proton side. Therefore, once the $N$=28 shell gap is quenched, the strong
quadrupole correlations amongst protons and neutrons will be ignited and lead to a novel
variety of the excitation modes. Indeed, various exotic phenomena such as the shape transition in Si
and S isotopes and the coexistence of various deformed states are theoretically suggested 
\cite{zpLi11,RE11,mK13}. 

In this paper, we report the low-lying quadrupole excitation modes in
$^{46}$Ar, $^{44}$S and $^{42}$Si generated by the strong quadrupole correlation between
protons and neutrons. To access the low-lying quadrupole modes of these isotopes, we apply
the Canonical-basis time-dependent Hartree-Fock-Bogoliubov (Cb-TDHFB) theory \cite{EN10} 
which has been successfully applied to the study of the dipole \cite{EN10, EN14} and 
quadrupole \cite{SL13} resonances of nuclei of wide mass range. 
The Cb-TDHFB can describe self-consistently the dynamical effects of pairing correlation which has 
a significant role to generate the low-lying quadrupole strength as reported in Ref. \cite{YG04,YYM06,LPD10,yH12,yH13}. 

The paper is organized as follows. 
In Sec.~\ref{sec:method}, 
we present the Cb-TDHFB equations, 
the choice of energy density functional
and the method to evaluate the strength function of quadrupole modes and $B$(E2). 
In Sec.~\ref{sec:results}, %at first 
the results of $^{48}$Ca and $^{46}$Ar obtained by using Skyrme SkI3 parameter set 
are presented, and the enhancement of the low-lying quadrupole mode in $^{46}$Ar is demonstrated. 
The interaction dependences of the results are also discussed. 
Furthermore, the low-lying states of other $N$=28 isotones ($^{44}$S, $^{42}$Si) are also investigated. 
Finally, Sec.~\ref{sec:concl} summarizes this work. 

\section{Formulation}
In this section, we introduce the Cb-TDHFB equation briefly, 
choice of interaction and the method to evaluate quadrupole strength function and $B$(E2). 
\label{sec:method} 
\subsection{Cb-TDHFB equation}
By assuming the diagonal form of pairing functional, 
the Cb-TDHFB equations \cite{EN10} are derived from 
the full TDHFB equation represented in the canonical basis 
$\{ \phi_l(t), \phi_{\bar l}(t)\}$ which diagonalize a density matrix.  
The Cb-TDHFB equations describe the time-evolution of the canonical pair 
$\{ \phi_l(t),\phi_{\bar l}(t) \}$, its occupation probability $\rho_l(t)$ and pair
probability $\kappa_l(t)$,  
\begin{eqnarray}
\label{Cb-TDHFB}
\
&i&\frac{\partial}{\partial t} \ket{\phi_l(t)} =
(h(t)-\eta_l(t))\ket{\phi_l(t)},\\ %\hspace{3mm} (\ {\rm for}\ \ \bar{l}\ ) \nonumber\\
&i&\frac{\partial}{\partial t} \ket{\phi_{\bar l}(t)} =
(h(t)-\eta_{\bar l}(t))\ket{\phi_{\bar l}(t)}, \nonumber\\
&i&\frac{d}{dt}\rho_l(t) =
\kappa_l(t) \Delta_l^{\ast}(t)
-\kappa_l^{\ast}(t) \Delta_l(t) , \nonumber\\
&i&\frac{d}{dt}\kappa_l(t) =
\left(
\eta_l(t)+\eta_{\bar l}(t)
\right) \kappa_l(t) + \Delta_l(t) \left( 2\rho_l(t) -1 \right) ,\nonumber
\end{eqnarray}
where $\eta_l(t)\equiv\bra{\phi_l(t)}h(t)\ket{\phi_l(t)}$, and the $h(t)$ and $\Delta_l(t)$
are the single-particle Hamiltonian and the gap energy, respectively. 
The numerical calculation was performed
in the three-dimensional Cartesian coordinate space representation. 
The canonical basis $\phi_l(\vect{r},\sigma; t) = \langle \vect{r},\sigma | \phi_l(t) \rangle$ 
with $\sigma=\pm 1/2$ is expressed in the space which is discretized 
in a square mesh of 1.0 fm inside of a sphere of radius 18 fm. 
We introduce the absorbing boundary condition (ABC) 
to eliminate unphysical modes by adding a complex potential $-i\xi(\vect{r})$ to 
the single-particle Hamiltonian $h(t)$ in accordance with Refs.~\cite{NY05,NY01}. 
Here $\xi(\vect{r})$ is given as, 
\begin{eqnarray}
\xi (\vect{r}) = 
\begin{cases}
\ 0, & (\mbox{for }\ 0 < r < R_0) \\
\ \xi_0 \dfrac{|\vect{r}|-R_0}{r_{\rm ABC}}, & (\mbox{for } R_0 < r < R)\\
\end{cases}
\end{eqnarray}
where $R$=$R_0$+$r_{\rm ABC}$=18 fm. 
We set the ABC effective range $r_{\rm ABC}$=6 fm and the depth $\xi_0$=3.75 MeV.

\subsection{The choice of energy density functional} 
We apply the Skyrme functionals to particle-hole ($ph$)-channel. 
Unless otherwise mentioned explicitly, we adopt the SkI3 parameter set~\cite{RF95} 
which, amongst the parameter sets we tested, 
most reasonably describes the proton and neutron single-particle levels in the vicinity of $^{48}$Ca. 
It is experimentally known that $N$=28 shell gap is approximately 4.8 MeV~\cite{lG06} and 
the proton $s_{1/2}$ is a few hundreds keV above $d_{3/2}$~\cite{DWK76,smB85}. 
To investigate the interaction dependence of the results, 
we also show the results obtained SkM$^*$ parameter set~\cite{jB82}, 
that yields smaller $N$=28 shell gap and different order of proton $s_{1/2}$ and $d_{3/2}$ shells. 

We use a schematic pairing functional form that was employed in Ref.~\cite{EN10}. 
The pairing energy $E_{\rm pair}$ 
and gap parameter $\Delta_l(t)$ are written as 
\begin{eqnarray}
E_{\rm pair} \equiv - \sum_{\tau=n,p} G_0^{\tau} 
\sum_{k,l > 0} f(\varepsilon_k^{\tau 0})f(\varepsilon_l^{\tau 0}) \kappa_k^{\tau}(t) \kappa_l^{\tau\ast}(t),
\end{eqnarray}
where, the cut-off function $f(\varepsilon)$ depends on 
the single-particle energy $\varepsilon_l^{\tau 0}$ in the ground state \cite{EN10,EN14}. 
In this work, we have tested several choice of the pairing strength $G_0^\tau$, 
that will be explained in the next section. 

\begin{table*}[t]
\caption{Properties of the single-particle levels of $^{48}$Ca and $^{46}$Ar around the
 chemical potential. $J$, $\varepsilon_{\rm sp}$ and $\rho_l$ denote their angular momenta, 
 energies and occupation probabilities, respectively. 
 When $\rho_l$ is an integer (``1" or ``0"), the nucleus has no superfluid phase in neutron or proton. 
 The single-particle states more than 7.5 MeV above the chemical potential are not shown, 
 due to the cutoff energy in the pairing channel \cite{EN10}.}
\label{tab:gs}
\begin{tabular}{ccccccccccccccc}\hline \hline
\multicolumn{7}{c}{$^{48}$Ca}& &\multicolumn{7}{c}{$^{46}$Ar} \\
\cline{1-7} \cline{9-15}
\multicolumn{3}{c}{neutron}&& \multicolumn{3}{c}{proton} && \multicolumn{3}{c}{neutron}&& \multicolumn{3}{c}{proton}\\
\cline{1-3} \cline{5-7}\cline{9-11} \cline{13-15}
$J$ & $\varepsilon_l$ &$\rho_l$&& $J$ & $\varepsilon_l$ &$\rho_l$&& $J$ & $\varepsilon_l$ &$\rho_l$&& $J$ & $\varepsilon_l$& $\rho_l$ \\
\cline{1-7} \cline{9-15}
$f_{5/2}$ &  -1.42, &0&&           &         & &&           &        & &&           &         &        \\
$p_{1/2}$ &  -3.00, &0&&           &         & && $p_{1/2}$ &  -2.10,&0&&           &         &        \\
$p_{3/2}$ &  -4.78, &0&&           &         & && $p_{3/2}$ &  -3.61,&0&&           &         &        \\
$f_{7/2}$ &  -9.78, &1&& $f_{7/2}$ &  -9.18, &0&& $f_{7/2}$ &  -8.09,&1&& $f_{7/2}$ &  -9.56, &  0.00  \\
$d_{3/2}$ & -19.05, &1&& $s_{1/2}$ & -16.45, &1&& $d_{3/2}$ & -17.39,&1&& $s_{1/2}$ & -17.11, &  0.68  \\
$s_{1/2}$ & -19.13, &1&& $d_{3/2}$ & -17.02, &1&& $s_{1/2}$ & -17.58,&1&& $d_{3/2}$ & -17.36, &  0.71  \\
$d_{5/2}$ & -25.28, &1&& $d_{5/2}$ & -23.42, &1&& $d_{5/2}$ & -23.62,&1&& $d_{5/2}$ & -23.76, &  0.97  \\ \hline 
\end{tabular}
\end{table*}

\subsection{Strength function and $B$(E2)}
In order to induce quadrupole responses, we add a weak instantaneous external field 
$V_{\rm ext}(\bm{r},t)=\eta\hat{F}_K(\bm r) \delta(t)$ to initial states of the time evolution. 
Here the quadrupole external field acting on proton, neutron, 
isoscalar (IS) and isovector (IV) channels are given as
$\hat{F}_{K} \equiv  (\frac{1\mp\tau_z}{2},1{\rm\ or\ }\tau_z)\otimes (r^2 Y_{2K} + r^2
Y_{2-K})/\sqrt{2(1+\delta_{K0})}$.  
The amplitude of the external field is so chosen to be a small number
$\eta=1 \sim 3\times10^{-3}$ fm$^{-2}$ to guarantee the linearity.
The strength function $S(E;\hat{F}_K)$ in each channel is obtained through the Fourier
transformation of the time dependent expectation value 
${\cal F}_K(t) \equiv \langle \Psi(t)| \hat{F}_K | \Psi(t) \rangle$:  
\begin{eqnarray}
\label{eq:str}
S(E;\hat{F}_K) &\equiv& \sum_{n} |\langle \Psi_n | \hat{F}_K | \Psi_0 \rangle |^2
 \delta(E_n - E) \nonumber \\ 
 &=& 
 \frac{-1}{\pi \eta} {\rm Im} \int_{0}^{\infty} \!\! \{ {\cal F}_K(t) - {\cal F}_K(0) \} 
 e^{i(E+i\Gamma/2)t} dt, \nonumber \\
\end{eqnarray}
where $| \Psi_0 \rangle$ and $| \Psi_n \rangle$ are the ground and excited states respectively, 
while $| \Psi (t) \rangle$ is a time-dependent many body wave function described by Cb-TDHFB equations. 
$\Gamma$ is a smoothing parameter set to 1 MeV. 
We also performed  unperturbed calculations in which $h(t)$ in Eq. (\ref{Cb-TDHFB}) is
replaced with the static single particle Hamiltonian $h(t=0)$ computed using the ground state density.
By comparing the results obtained by the fully self-consistent and unperturbed calculations, 
we investigate the effects of the residual interaction on the collectivity of the low-lying states. 

To analyze the peak structure, we fit the $S(E;\hat{F}_K)$
below 10 MeV by the sum of Lorentzian ${\frak L}_k(E;\hat{F}_{K})$:  
\begin{eqnarray} 
\label{eq:fit}
S(E;\hat{F}_K) 
&\simeq& \sum_{k} {\frak L}_k(E;\hat{F}_{K}) \nonumber \\ 
&\equiv& \sum_{k} \frac{a_k^{\tau} (\Gamma/2)^2}{(E-E(2_k^+)^2 + (\Gamma/2)^2},
\end{eqnarray}
where $k$ is a label of 2$^+$ state 
and $\Gamma$ corresponds to the smoothing parameter in Eq.(\ref{eq:str}). 
The reduced transition probabilities in the proton and neutron channels, 
that are denoted as $B_k$(E2$\uparrow$) and $B_k$(N2$\uparrow$) 
are evaluated by integrating ${\frak L}_k(E;\hat{F}_K)$ for each state, 
\begin{eqnarray} 
\label{eq:be2}
B_k({\rm E2}\!\uparrow \ {\rm or}\ {\rm N2}\!\uparrow) &\equiv& 
\! \sum_{K} |\langle 2_k^+| \hat{Q}_{2K}\frac{1\mp \tau_z}{2} |0^+ \rangle|^2  \nonumber
\\ 
&\simeq& 5\! \int_0^{\infty} \! {\frak L}_k(E;\hat{F}_{K=0})\ dE.
\end{eqnarray}
The approximation in bottom line of Eq.(\ref{eq:be2}) is reasonable for spherical nuclei
such as $^{48}$Ca and $^{46}$Ar, but gets worse for deformed nuclei $^{44}$S and $^{42}$Si.

We mention spurious modes in the present calculations. 
In our scheme, spurious modes (pairing rotation, spatial rotation, etc.)
come up at zero-energy in principle \cite{EN10}. 
By the accurate time-evolution using the 4th order expansion of Cb-TDHFB 
to conserve particle number, 
and by preparing the ground state wave function with 
the expectation value of center-of-mass less than 10$^{-12}$ fm.  
These spurious modes are greatly reduced and almost invisible. 

\section{Results and discussions}
\label{sec:results}
\subsection{$^{48}$Ca and $^{46}$Ar}
\label{ssec:48Ca46Ar}
\begin{figure*}[t]
 \begin{center}
  \includegraphics[keepaspectratio,width=7cm, angle=-90]{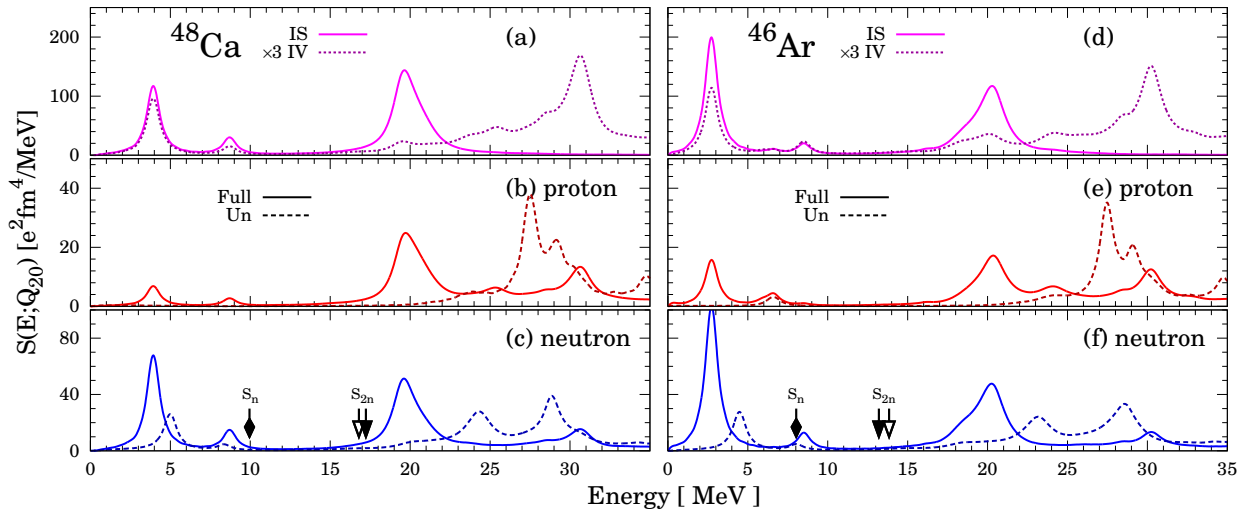}
  \caption{(Color on-line) Strength functions of quadrupole vibrational modes of
  $^{48}$Ca and $^{46}$Ar. The strength functions in the IS (solid line) and
  IV (dotted line) channels are shown in the panels (a) and (d), while those in the
  proton and neutron channels are shown in (b), (e) and (c), (f). 
  The solid and dashed lines in the panels (b), (c), (e) and (f) 
  compare the fully self-consistent and unperturbed results. 
  The filled arrows show the experimental one- and two-neutron
  separation energies \cite{mW12}, while the open arrows show 
  the calculated two-neutron separation energies.}
  \label{fig:48Ca46ArQ}
 \end{center}
\end{figure*}

First, we examine the quadrupole responses of double-closed-shell nucleus $^{48}$Ca 
which is spherical and has no superfluid phase. 
The calculated neutron $pf$-shell gap 5.0 MeV,  
the order of proton $s_{1/2}$ and $d_{3/2}$ and their energy gap 0.57 MeV are consistent with 
the experimental data in the vicinity of $^{48}$Ca~\cite{DWK76,smB85,lG06}.
The strength functions in the IS and IV channels shown in Fig.~\ref{fig:48Ca46ArQ} (a) 
have two peaks at 3.9 and 8.7 MeV in addition to the IS giant quadrupole resonance (GQR) 
at 20 MeV and the IVGQR having broad distribution around 30 MeV. 
The properties of these two peaks below 10 MeV become clear by
comparing the results in the proton and neutron channels (Fig.~\ref{fig:48Ca46ArQ} (b) and (c)). 
Their strengths in the neutron channel are much larger than those in the proton channel 
showing the dominance of neutron excitation and 
their energies approximately agree with the neutron shell gaps between $f_{7/2}-p_{3/2}$ (5.0 MeV)
and $f_{7/2}-f_{5/2}$ (8.4 MeV). 
Now we focus on the lowest peak to examine how the residual interaction affects the low-lying 
quadrupole modes by comparing the unperturbed and fully self-consistent results. 
The lowest peak at 5.0 MeV ($f_{7/2} \to p_{3/2}$) in the unperturbed neutron channel is 
lowered by 1.1 MeV and its strength is increased by a factor of 2.5 in the fully self-consistent results
as shown in Tab.~\ref{tab:E2-1}. 
The residual interaction between protons and neutrons induces 
a weak strength in the proton channel at the same energy 
where no strength exists in the unperturbed result, 
and hence this proton small peak cannot be attributed to the proton single-particle excitations. 
The evaluated first peak energy is $E$(2$^+$)=3.9 MeV 
and $B$(E2)=52 $e^2$fm$^4 \simeq$1.0 W.U. (Weisskopf unit) listed in Tab.~\ref{tab:E2-1} 
reasonably agree with those of the observed 2$^+$ state, 
$E$(2$^+$)=3.83 MeV and $B$(E2)=95$\pm$32 $e^2$fm$^4$=1.8$\pm$0.6 in W.U. \cite{RNT01}. 
Although the observed and calculated $B$(E2) strength are not as large as to be ``collective", 
this result demonstrates that the residual interaction between protons and neutrons potentially 
induces proton collectivity even in the double-closed-shell nuclei. 
Naturally, we expect that this induction mechanism strongly depends on the size of $N$=28 
shell gap and the shell ordering and energy gap of proton $sd$-shell, 
and we will see the proton collectivity is reinforced in neutron deficient $N$=28 isotones in the following. 

Now, we discuss the low-lying 2$^+$ states of $^{46}$Ar. % based on the results in $^{48}$Ca. 
Here the proton pairing strength $G_0^p$ is chosen so that the pairing gap reproduces
the empirical value $\Delta^p$=2.7 MeV evaluated 
by three-points formula\footnote{$\Delta^p(N,Z)=(2B(N,Z)-B(N,Z-1)-B(N,Z+1))/2$, 
where $B(N,Z)$ is a binding energy.}, 
while that for neutron is set to zero which is consistent with the HFB calculations with Gogny force \cite{PGF00}. 
Other choice of pairing strength and the dependence of the results on the choice 
will be discussed in the next subsection. 
With this choice of $G_0^p$, the ground state of $^{46}$Ar is spherical. % but when the superfluid phase appears in proton.
If the proton pairing is switched-off the ground state is oblately deformed ($\beta$=$-$0.18), 
which implies the weaker magicity of $N$=28 in $^{46}$Ar than $^{48}$Ca. 
Indeed, the calculated $N$=28 shell gap (see Tab.~\ref{tab:gs}) is slightly reduced in $^{46}$Ar (4.5 MeV) 
compared to that in $^{48}$Ca (5.0 MeV), 
which is consistent with observed gaps of $^{46}$Ar (4.4 MeV) and $^{48}$Ca (4.8 MeV) \cite{lG06}.
The unperturbed strengths in the proton and neutron channels in the Fig.~\ref{fig:48Ca46ArQ} 
(e) and (f) are quite similar to those of $^{48}$Ca except for minor differences; 
(1) the reduction of the lowest peak energy 
in the neutron channel due to the quenching of the $N$=28 shell gap, 
and (2) the small peak located around 6.6 MeV in the proton channel 
that is generated by the proton hole-states in $sd$-shell.  

In the fully self-consistent results, 
strong peaks emerge around 2.7 MeV in all channels as shown in Fig.~\ref{fig:48Ca46ArQ} (d)-(f). 
The neutron strength function looks similar to that of $^{48}$Ca except for the very pronounced lowest peak. 
Similar to $^{48}$Ca, the corresponding proton peak at the same energy is also induced 
due to the residual interaction. 
However, it is notable that energy of the lowest peak is considerably lowered 
and its strength is much enhanced than $^{48}$Ca, 
and their values are $B$(E2$\uparrow$)=114 $e^2$fm$^4$ 
and $B$(N2$\uparrow$)=748 $e^2$fm$^4$. 
\begin{table}[h]
\caption{Peak position $E(2^+;\hat{F}_{K=0})$ [MeV] and 
 the evaluated transition strengths $B$(N2$\uparrow$) and $B$(E2$\uparrow$) 
 [$e^2 \rm fm^4$] are also shown for $^{48}$Ca and $^{46}$Ar. %in Sec.~\ref{ssec:48Ca46Ar}
 The height $a_k^{\tau}$ [$e^2$ fm$^4$/MeV] of Lorentzian obtained by 
 fitting the strength functions below 10 MeV (see Eq.(\ref{eq:fit}) and (\ref{eq:be2})).} 
\label{tab:E2-1}
\begin{tabular}[t]{p{10mm}cccrr} \hline \hline 
 &\ \ $E(2^+)$,\ & $B$(E2$\uparrow$), & $B$(N2$\uparrow$), & $a_k^{\rm p}$, & $a_k^{\rm n}$ \\ \hline 
$^{48}$Ca  & 3.9 &  52   & 510 &  6.9 &  67.8 \\
           & 8.7 &  21   & 110 &  2.8 &  14.3 \\
$^{46}$Ar  & 2.7 & 114   & 748 & 15.5 & 101.3 \\
           & 6.6 &  36   &  -- &  4.7 &   --  \\
           & 8.5 &  --   &  94 &  --  &  12.2 \\
\hline
\end{tabular} 
\end{table}
Experimentally, small and large values of $B$(E2;$0_1^+\to2_1^+$) at 1.55 MeV have been reported; 
in the Coulomb-excitation study 
$B$(E2$\uparrow$)=196$\pm$39 (218$\pm$31) $e^2$fm$^4$=4.0$\pm$0.8(4.4$\pm$0.6) W.U. \cite{hS96,aG03}, 
and in the Life-time measurement $B$(E2$\uparrow$)=570$_{-160}^{+335} e^2$fm$^4$ 
=11.6$_{-3.3}^{+7.8}$ W.U. \cite{dM10}. 
In both experiments, the E(2$^+$) is decreased and the $B$(E2) is enhanced compared to $^{48}$Ca, 
although they have large discrepancy \cite{dM10,rW12}. 

In this section, we have focused on the low-lying quadrupole mode of the nuclei 
which have a neutron magic number $N$=28 and 
their shape are spherical {\it i.e.} $^{48}$Ca and $^{46}$Ar. 
The hole-state in $sd$-shell is a very important ``seed" to generate a strong lowest 2$^+$ state
and to induce the $pn$-quadrupole correlation which does not appear in the double-closed-shell nucleus. 
The keys that the mode to get a stronger collectivity in $^{46}$Ar, 
will be due to the proton hole-states and the neutron $pf$-shell quenching, 
while working the pairing correlation. 
To understand the mechanism of the 2$^+$ state, 
we investigate the interaction dependences in $ph$- and $pp$($hh$)-channel, namely 
structure and correlation dependences, in the next subsection. 

\subsection{Interaction dependence of $^{46}$Ar}
\label{sec:intD}
\begin{table*}[t]
\caption{Same as Tab.~\ref{tab:gs}, but for $^{46}$Ar$^{\rm I, II, III}$.}
\label{tab:gs2}
\begin{tabular}{cccccccccccccccccccc}\hline\hline
 &\multicolumn{5}{c}{$^{46}$Ar$^{\rm I}$ (w/SkM$^*$)}& &\multicolumn{5}{c}{$^{46}$Ar$^{\rm II}$ (w/$\Delta^n\neq$ 0)} 
 &&\multicolumn{6}{r}{$^{46}$Ar$^{\rm III}$ (w/weak $_{\Delta^p\neq0}^{\Delta^n\neq0}$)}\\
\cline{2-6} \cline{8-12} \cline{14-20}
  & \multicolumn{2}{c}{neutron} && \multicolumn{2}{c}{proton} && \multicolumn{2}{c}{neutron}&& \multicolumn{2}{c}{proton}
  &&\multicolumn{3}{c}{neutron} && \multicolumn{3}{c}{proton} \\
 \cline{2-3} \cline{5-6} \cline{8-9} \cline{11-12} \cline{14-16} \cline{18-20}
  $J$& $\varepsilon_l$ & $\rho_l$ && $\varepsilon_l$ & $\rho_l$ & & $\varepsilon_l$ & $\rho_l$ & & $\varepsilon_l$& $\rho_l$
&&$J$& $\varepsilon_l$ & $\rho_l$ && $J$& $\varepsilon_l$ & $\rho_l$\\ 
\cline{1-12} \cline{14-20} 
$p_{1/2}$ &  -3.06  &0&&        &      & &  -2.05, &  0.08  &&        &      && $p_{1/2}$ &  -2.07  &0.04&&           &         &     \\
$p_{3/2}$ &  -4.95  &0&&        &      & &  -3.63, &  0.20  &&        &      && $p_{3/2}$ &  -3.61  &0.11&&           &         &     \\
$f_{7/2}$ &  -8.26, &1&& -10.09 & 0.03 & &  -8.02, &  0.89  &&        &      && $f_{7/2}$ &  -8.05  &0.94&&           &         &     \\
$d_{3/2}$ & -12.83, &1&& -15.36 & 0.61 & & -17.08, &  0.99  && -16.89 & 0.69 && $d_{3/2}$ & -17.22  &1.00&& $s_{1/2}$ & -17.11  &0.69 \\
$s_{1/2}$ & -15.22, &1&& -16.57 & 0.78 & & -17.57, &  0.99  && -17.07 & 0.71 && $s_{1/2}$ & -17.57  &1.00&& $d_{3/2}$ & -17.12  &0.69 \\
$d_{5/2}$ & -19.34, &1&& -21.81 & 0.97 & & -23.46, &  1.00  && -23.39 & 0.97 && $d_{5/2}$ & -23.54, &1.00&& $d_{5/2}$ & -23.58, &0.98 \\ 
\hline 
\end{tabular}
\end{table*}
To investigate how the enhancement of the low-lying strength depends on 
the underlying nuclear structure, 
we tested several different combinations of the energy density functional in 
$ph$- and $pp$($hh$)-channels. 
In any cases we tested, the low-lying quadrupole collectivity in $^{46}$Ar 
is much more enhanced than in $^{48}$Ca, but its magnitude and distribution depend on 
the single-particle structure and pairing strength. 
To illustrate it, we discuss three cases which are denoted as 
$^{46}$Ar$^{\rm I}$, $^{46}$Ar$^{\rm II}$ and $^{46}$Ar$^{\rm III}$ in the following. 

\begin{figure}[h]
  \begin{center}
   \includegraphics[clip, keepaspectratio, width=5cm, angle=-90]{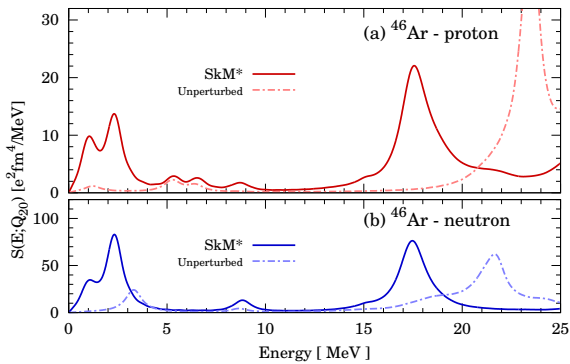}
   \caption{(Color on-line) Strength functions in proton (red) and neutron (blue) channels for
   $^{46}$Ar using SkM$^*$. 
   Fully self-consistent (thin solid line) and unperturbed (thin chain line) results are shown. }
   \label{fig:2}
   \end{center}
\end{figure}

The first case $^{46}$Ar$^{\rm I}$ is calculated by using SkM$^*$ 
in $ph$-channel instead of SkI3, and the pairing strengths are determined by 
the same manner with that explained in Sec.~\ref{ssec:48Ca46Ar}, 
{\it i.e.} the strength of proton $G_0^p$ is so chosen to reproduce 
the empirical pairing gap, while that of neutron is vanished. 
This choice of the energy functional yields the proton and neutron 
single-particle levels listed in Tab.~\ref{tab:gs2} where the $N$=28 shell gap 
(3.31 MeV) is smaller than those of SkI3 and observations, and 
the order of proton $d_{3/2}$ and $s_{1/2}$ is reverted 
in contradiction to the observations. 
Figure~\ref{fig:2} shows the strength functions of $^{46}$Ar$^{\rm I}$ 
in (a) proton and (b) neutron channels, 
which are obtained in the fully self-consistent (solid lines) 
and the unperturbed (thin chain lines) calculations. 
As clearly seen, the enhancement of the low-lying collectivity similar to the results 
shown in Fig.~\ref{fig:48Ca46ArQ} is confirmed. 
However, there are several differences between the SkI3 and SkM$^*$ results; 
first, the low-lying peak is split into two, and second, 
the amount of the low-lying strength is increased. 
The origin of the splitting may be attributed to the proton excitation from 
$s_{1/2}\to d_{3/2}$. As already mentioned, SkM$^*$ yields different ordering of 
the proton $sd$-shell, $d_{3/2}$ is 1.2 MeV above $s_{1/2}$, and hence, 
the corresponding proton excitation can be seen as a small peak at 1.2 MeV 
in the unperturbed proton strength function. 
By switching on the residual interaction, the proton excitation couples 
to the neutron excitations and grows to the peak at 1.1 MeV in the full-calculation. 
On the other hand, the peak at 2.3 MeV has similar nature to that of SkI3 result, {\it i.e.}, 
it is dominated by the neutron excitations across $N$=28 shell gap 
and the proton excitation is induced by the residual interaction. 
The relatively reduced $N$=28 shell gap in SkM$^*$ dearly increases the amount of 
the strengths below 5 MeV (S$_{\rm E2}^{\rm 5 MeV}\equiv 5 \int_0^{5} S(E;E2)dE$) 
from 109 $e^2$fm$^4$ (SkI3) to 139 $e^2$fm$^4$ (SkM$^*$). 
The other visible differences are the small peaks around 7 MeV in proton channel, 
which originate in the proton $d_{5/2}\to s_{1/2}$ and $d_{3/2}$ excitations and 
corresponds to the splitting of $s_{1/2}$ and $d_{3/2}$. 
The GQR locates at 17.5 MeV in lower than that of SkI3, which has no drastic change. 

\begin{figure}[h]
  \begin{center}
   \includegraphics[clip, keepaspectratio, width=5cm, angle=-90]{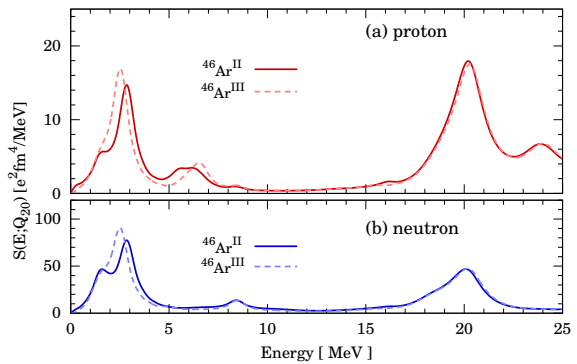}
   \caption{(Color on-line) Strength functions in proton (red) and neutron (blue) channels for 
   $^{46}$Ar$^{\rm II}$ (solid line) and $^{46}$Ar$^{\rm III}$ (dashed line), 
   which are obtained by the fully self-consistent calculation.}
   \label{fig:3}
   \end{center}
\end{figure}

It is well-known that the pairing correlation has 
a significant role to form the low-lying quadrupole modes as reported in Ref.\cite{YG04,YYM06,LPD10,yH12,yH13}.
Therefore, we now discuss two cases ($^{46}$Ar$^{\rm II, III}$) 
in order to investigate the pairing effects to low-lying 2$^+$ strength. 
The case $^{46}$Ar$^{\rm II}$ is calculated by using SkI3 and the pairing strength $G_0^\tau$ 
are determined to reproduce empirical $\Delta^p$ and $\Delta^n$ simultaneously, 
in short, the pairing correlation in neutron channel is turned-on. 
In the case of $^{46}$Ar$^{\rm III}$, % has also superfluid phases in both channels, but 
its pairing-strengths are weakened to 85\% of those of $^{46}$Ar$^{\rm II}$. 
The calculated single-particle levels in Tab.~\ref{tab:gs2} show that 
the pairing in neutron channel changes the occupation probability of each orbit, 
but does not strongly affect the single-particle energies. 
The distributions of the strength functions shown in Fig.~\ref{fig:3} 
are not drastically changed by switching on the neutron pairing, 
except for a shoulder around 1.5 MeV which appears 
due to the partial occupation of neutron $p$-orbits. 
The major difference brought about by pairing correlation 
is the further enhancement of the low-lying collective mode. 
Here we check it by looking at the sum of the E2 strengths below 5 MeV;
S$_{\rm E2}^{\rm 5 MeV}$($^{46}$Ar, $^{46}$Ar$^{\rm II}$, $^{46}$Ar$^{\rm III}$)=
109, 128 and 131 $e^2$fm$^4$ are obtained. 
These results indicate that 
the pairing correlation tends to enhance the low-lying collectivity, 
but its magnitude is sensitive to the pairing strength.  

Here, we also comment on the comparison between the present result and experimental data. 
The Life-time measurement \cite{dM10} reports 
about three times larger value of $B$(E2)$\sim$570 $e^2$fm$^4$
than that reported by Coulomb-excitation experiments \cite{hS96,aG03}, 
$B$(E2)$\sim$196(218) $e^2$fm$^4$. 
In all cases we tested, the low-lying strengths are less than 200 $e^2$fm$^4$, 
and hence supports the smaller value of $B$(E2) reported by Coulomb excitation. 

As summary of this section, for all cases we tested, 
it is found that the low-lying collectivity is enhanced. 
It is also noted that the distribution and magnitude depend on 
the single-particle structure and pairing strength. 

\subsection{$^{42}$Si and $^{44}$S}
\begin{figure}[h]
  \begin{center}
   \includegraphics[clip, keepaspectratio, width=6cm, angle=-90]{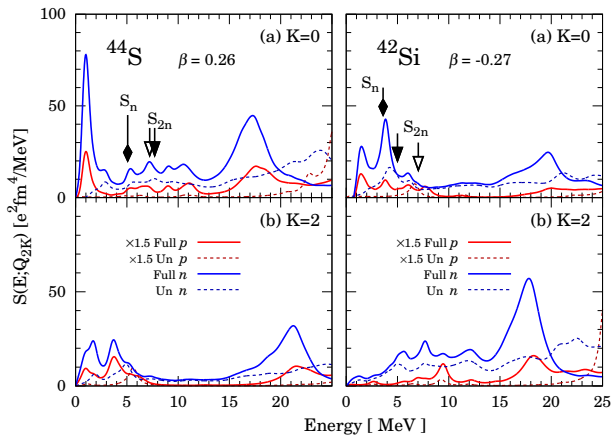}
   \caption{Strength functions in the proton and neutron channels for 
   $^{44}$S and $^{42}$Si. $K=0$ and 2 modes are shown in panels (a), (c) and (b), (d), respectively. 
   The experimental separation energies S$_n$, S$_{2n}$ are taken from Ref. \cite{mW12}.}
   \label{fig:44S42SiQ2}
   \end{center}
\end{figure}

The same mechanism also applies to other $N$=28 isotones $^{44}$S and $^{42}$Si 
which are experimentally known to have deformed shape 
and low-lying 2$^+$ states \cite{tG97,fS00,dS02,sG04,jF06,bJ07,mZ09,cF10,dsG11,jF05,bB07,lG08,sT12}. 
In the present calculation, 
the pairing strengths in both channels are chosen to reproduce empirical gap energies. 
$^{44}$S has prolate shape ($\beta$=0.26) with 
$\Delta^p$=2.37 MeV and $\Delta^n$=1.12 MeV, 
while $^{42}$Si has oblate shape ($\beta$=$-$0.27) with 
$\Delta^p$=2.70 MeV and $\Delta^n$=1.04 MeV, respectively. 
It is noted that the pattern of the deformed shape of the $N$=28 isotone is consistent 
with other theoretical results \cite{RD99,KSM11,zpLi11,RE11,UO12,mK13}. 
Figure \ref{fig:44S42SiQ2} shows their strength functions. 
Deformation of these nuclei splits the strength functions in the $K=0$ and 2 modes,  
and makes their strength distributions even more complicated than those in previous sections. 
However, we can still identify very low-lying peaks which correspond to a couple of 2$^+$ 
states with enhanced collectivity. 
It is noted that, for example, three 2$^+$ states are observed in $^{44}$S \cite{dsG11} 
and our results may be associated with some of them.

We also estimated 2$^+$ energies and $B$(E2$\uparrow$) of $^{44}$S and $^{42}$Si by
using Eq.(\ref{eq:fit}) and (\ref{eq:be2}), although quantitative comparison with
observation is not possible because of deformation of these nuclei. 
The first peak of $^{44}$S at 1.0 MeV has a value of $B$(E2$\uparrow)$=103 $e^2$fm$^4$ and 
that of $^{42}$Si at 1.4 MeV has $B$(E2$\uparrow)$=47 $e^2$fm$^4$. 
Similar to $^{46}$Ar, the energies of 2$^+$ states have shifted to lower energy, 
and their strengths are significantly increased compared to $^{48}$Ca. 
Indeed in the experimental study of $^{44}$S \cite{dS02,sG04,mZ09,cF10}, 
a couple of low-lying 2$^+$ states and the enhancements of $B$(E2) are reported, 
however, the observed $B$(E2) is 310$\pm$90 $e^2$fm$^4$ which is larger than our present result. 
This discrepancy could be attributed to the rotational mode 
caused by deformation of the ground band as reported by experiments \cite{tG97,sT12}. 
It is also noted the linear response method occasionally underestimates the observed $B$(E2) 
as reported by Scamps and Lacroix \cite{SL13} which could be another reason of the disagreement 
in $^{44}$S and $^{42}$Si.

\section{Conclusion}
\label{sec:concl}
We have investigated the low-lying quadrupole vibrational modes of $N$=28 isotones 
($^{48}$Ca, $^{46}$Ar, $^{44}$S, $^{42}$Si) by using Cb-TDHFB theory. 
The proton low-lying 2$^+$ states of $^{48}$Ca have neutron single-particle nature, 
which is found from the comparison fully self-consistent and unperturbed results. 
Our calculation well reproduce not only the single-particle levels of ground state but 
also the first 2$^+$ state and $B$(E2) of $^{48}$Ca. 
We discuss the mechanism of low-lying 2$^+$ excitation of other isotones, 
based on the $^{48}$Ca results. 
The 2$^+$ states of $^{46}$Ar have a similar mechanism of $^{48}$Ca, 
and the lowest peak indicates some collectivity which is triggered by proton hole-states. 
Furthermore, we investigate the behavior of 2$^+$ strength functions under the change of 
interaction ($ph$-, $pp$($hh$)-channel) 
for $^{46}$Ar. 
From the investigation, we found that the peak position and the collectivity of the 2$^+$ states 
strongly depend on the relation amongst the neutron $pf$-shell gap, 
the level-spacing in proton $sd$-shell and the pairing strength. 
Our studies argue that the low-energy 2$^+$ state bringing a strong collectivity 
can be generated by these three pieces, also in spherical system. 
Additionally, we found that the pairing correlation 
does not always enhance the collectivity of low-lying 2$^+$ states.
In this study, we obtain prolate and oblate deformed ground states for $^{44}$S and $^{42}$Si. 
The multiple 2$^+$ states which appear in deformed nuclei 
can be applied the argued mechanism, 
although it could not explain the lowest 2$^+$ state of deformed nuclei. 
The mechanism of 2$^+$ excited state can be induced near the region to violate ordinary magicity. 
We will apply the mechanism to other candidates, in future work.

\section*{Acknowledgment} 
We would like to thanks Prof. T. Nakatsukasa for giving us the useful comments and advice. 
Computational resources were partially provided 
by High Performance Computing system at Research Center for Nuclear Physics, Osaka University.

\end{document}